\documentclass[%
reprint,
longbibliography,
 amsmath,amssymb,
 aps,
]{revtex4-1}

\usepackage{graphicx}
\usepackage{dcolumn}
\usepackage{bm}
\usepackage{amsmath}
\usepackage{color}

\begin{document}


\title{Gradiometry-free noise suppression in RF atomic magnetometers}

\author{V.~Gerginov}
\email{Vladislav.Gerginov@colorado.edu}
\affiliation{Associate of the National Institute of Standards and Technology, Boulder, Colorado 80305, USA}%
\affiliation{Department of Physics, University of Colorado, Boulder, Colorado 80309, USA}

\date{\today}

\begin{abstract}
We demonstrate the sensitivity of a sensor based on an optically-pumped radio-frequency (RF) atomic magnetometer to the polarization state of the detected RF magnetic field, and measure $>$36\,dB difference in amplitude sensitivity for opposite circular field polarizations. This sensitivity could be used to create novel sensors that would allow signal detection while suppressing the ambient noise, regardless of the distance between the sources of signal and noise, in contrast to traditional gradiometry configurations. Additionally, such sensor will be orientation-sensitive, as the phase of the detected signal is shown to depend on the angle between the sensor's detection axis and the direction to the transmitter. 
\end{abstract}

\pacs{Valid PACS appear here}
\maketitle


\section{\label{sec:intro} Introduction}

Optically-pumped atomic magnetometers (OPMs) reach impressive sensitivities at the sub-femtotesla levels when measuring DC \cite{ShengLiDuralEtAl2013} or RF \cite{LeeSauerSeltzerEtAl2006} magnetic fields, and compete with state-of-the-art detectors based on Superconducting Quantum Interference Devices (SQUIDs) \cite{Clarke2010}. These devices are typically used in heavily shielded environments, for example in multilayer magnetically shielded rooms, or rely on active coils for ambient noise suppression. In an unshielded environment, the performance of the OPMs operating in the DC regime is degraded by several orders of magnitude due to the ambient magnetic field noise \cite{SeltzerRomalis2004}. Such situations can be remedied to some extent by averaging or by use of gradiometry techniques \cite{SlackVanceLynch1967}. The situation is better for the case of RF atomic magnetometers because the ambient noise, caused either by geomagnetic, atmospheric, or urban activities, is outside the bandwidth of interest \cite{Constable2016}. Nevertheless, noise suppression techniques are still required \cite{LeeSauerSeltzerEtAl2006,CooperPrescottLeeEtAl2018}.

The ultimate intrinsic sensitivity of the state-of-the-art magnetic field sensors is reached using two or more OPMs or OPM channels. The noise suppression techniques rely on the different distance to the signal and noise sources compared to the gradiometer baseline (the distance between the sensors). Early examples of distant source detection include magnetic anomaly detection and exploration of the Earth's magnetic field \cite{SlackVanceLynch1967}, as well as magnetic fields in space \cite{SlocumReilly1963}, while detection of local signal sources includes measurements of faint biomagnetic signals \cite{LivanovKoslovSinelnikovaEtAl1981}. 

In this work, we develop a two-channel optically-pumped atomic RF magnetometer capable of detecting low-frequency (below 1\,MHz) magnetic field signals. The quantum properties of the RF magnetometer are explored to construct balanced detection schemes for measurement or suppression of ambient noise. The schemes do not rely on gradiometer methods and are expected to be largely immune to the relative distance between the signal and noise sources. The proposed schemes rely on the quantum-mechanical laws of angular momentum conservation. They exploit the sensitivity of an RF atomic magnetometer to the polarization state of an RF magnetic signal \cite{SavukovSeltzerRomalisEtAl2005, OidaItoKamadaEtAl2012} in a novel way that is demonstrated theoretically and experimentally in this work, achieving $>$36\,dB rejection of the unwanted circular polarization of the RF field. A quantum receiver that includes ambient noise suppression, based on the results of this work, is proposed. This work mainly focuses on the application of the receiver for communication using artificial low-frequency magnetic signals in the presence of strong absorption \cite{GerginovSilvaHowe2017} and ambient (geomagnetic, atmospheric, urban) magnetic field noise, but it might be possible to use some of the results for different applications such as low-field NMR spectroscopy \cite{SavukovRomalis2005, LeeSauerSeltzerEtAl2006} and magnetic induction imaging \cite{DeansMarmugiHussainEtAl2016}.

\section{\label{sec:setup}Two-channel RF magnetometer}

\begin{figure}
\includegraphics[width=\columnwidth]{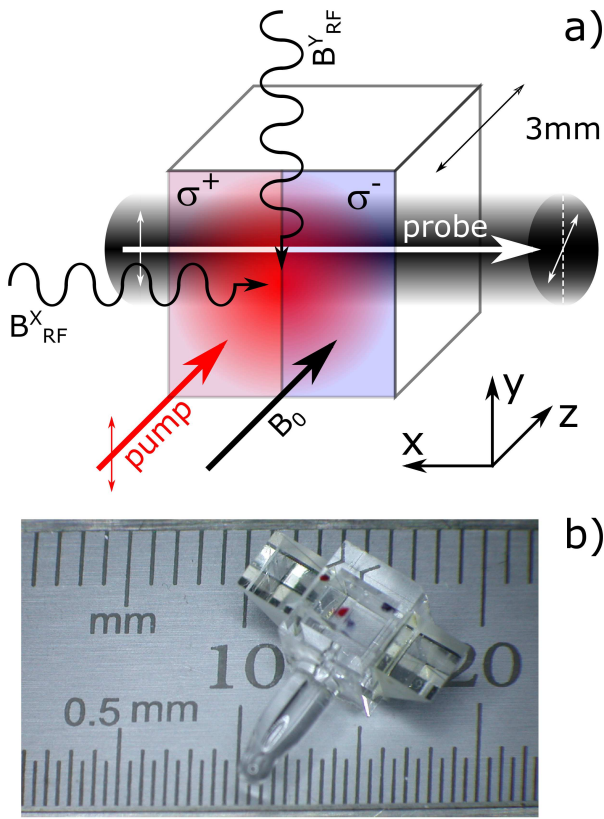}
\caption{\label{fig:Setup} a) Diagram of the two-channel optically-pumped RF magnetometer. b) Photograph of the actual optical cell assembly. The two right-angle prisms (not shown in a)) steer the probe beam direction before and after it passes the vapor cell to be parallel to the pump beam for convenience.}
\end{figure}

To study the sensitivity of the RF magnetometer to the polarization of the low-frequency magnetic fields, a two-channel RF magnetometer with an active volume of 27\,mm$^3$ was constructed. Figure\,\ref{fig:Setup} a) shows the optical cell assembly of the RF magnetometer. The cubic ($3 \times 3 \times 3$\,mm$^3$ internal volume) pyrex vapor cell (Triad Technology, Inc. \cite{disclaimer}) contains enriched $^{87}$Rb and 500\,Torr nitrogen as a buffer gas. The thickness of the cell's optical-quality windows is 0.5\,mm. Two $\lambda/4$ waveplates ($2 \times 2 \times 0.2$\,mm) are epoxied side-by-side on the front window (in the $xy$-plane) of the cell. The slow axes of the waveplates are orthogonal, and at $\pm45^{\circ}$ with respect to the $y$-axis. 

A linearly-polarized 4\,mm diameter pump laser beam at $\lambda=794.99(1)$\,nm (on resonance with the $6s$\,$^2S_{1/2} \rightarrow 6p$\,$^2P_{1/2}$ pressure-shifted atomic transition) propagating along the $z$-axis passes through the waveplates, creating two separate sections of the cell with pump light of opposite circular polarizations. The reduced $^{87}$Rb diffusion due to the buffer gas spatially separates the atoms pumped with light of opposite circular polarization. 

A linearly-polarized probe beam at $\lambda=780.12(1)$\,nm (blue-detuned from the $6s$\,$^2S_{1/2} \rightarrow 6p$\,$^2P_{3/2}$ pressure-shifted atomic transition by $>$50\,GHz) propagates through the cell along the negative $x$-axis. The incoming probe beam initially propagates along the $z$-axis and is directed in the vapor cell with an aluminum-coated $3$\,mm prism (see Fig.\,\ref{fig:Setup} b)). After the cell a second identical prism sends the probe beam in the negative $z$-direction through a polarimeter for balanced detection. The linear polarization of the probe beam experiences optical rotation in the presence of an atomic polarization precessing in the polarization plane of the probe beam \cite{BudkerRomalis2007}. 

The optical cell assembly shown in Fig.\,\ref{fig:Setup} b) is mounted in a 50\,mm long, 26\,mm diameter cylindrical oven made of machineable glass-ceramic. The oven accommodates the optical cell assembly and provides optical access for the pump and probe beams. It also allows the cell to be heated to $\sim$100\,$^{\circ}$C with a non-magnetic resistive heater driven by a current from a DC current. The oven is mounted in the center of a 3D printed coil assembly that is used to generate the static bias magnetic field $B_0$ parallel to the $z$-axis, and the two mutually orthogonal linearly-polarized RF magnetic fields $B^x_{RF}\left(t\right)$ and $B^y_{RF}\left(t\right)$ that are perpendicular to the bias field direction. The 3D coil assembly is housed in a three-layer magnetic shield chamber. 

The bias field $B_0$ is created by a pair of $100\times100$\,mm square coils in Helmholtz configuration by passing $\sim$40\,mA of low-noise DC current through them in series. The RF magnetic fields $B^x_{RF}\left(t\right)$ and $B^y_{RF}\left(t\right)$ are created by two RF coil pairs identical to the bias field coil pair. The current used to drive the RF magnetic fields is generated by an ac voltage applied across 200\,kOhm resistor connected in series with the pair of RF field coils. The voltage is generated by a two-channel function generator that allows precise phase control of the $B^x_{RF}\left(t\right)$ and $B^y_{RF}\left(t\right)$ signals.

\section{\label{sec:tune} Magnetometer response as a function of the bias magnetic field}

\begin{figure}
\includegraphics[width=\columnwidth]{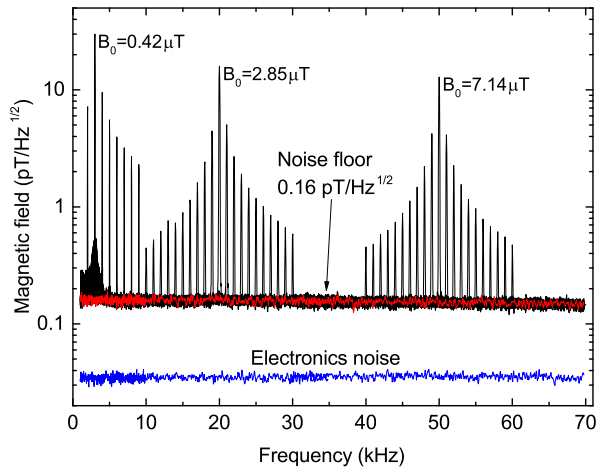}
\caption{\label{fig:Tuning} Tuning of the RF magnetometer with the value of the bias field $B_0$. The black traces are spectrum analyzer data with an applied RF magnetic signal of different frequency. The red curve (overlapping with the spectrum analyzer data away from the RF signal) shows the noise of the probe light in the absence of pump laser light. The blue trace shows the residual polarimeter electronics noise in the absence of light. The discrete signal peaks are separated by 1\,kHz, the steps of the $B^x_{RF}(t)$ signal.}
\end{figure}

In this Section we present results demonstrating the RF magnetometer tunability and noise floor. The measurements were performed before the quarter waveplates were installed, allowing the entire cell active volume of 27\,mm$^3$ to be used. The magnetometer bias field $B_0$ was chosen to correspond to three discrete Larmor precession frequencies (3, 20 and 50\,kHz, corresponding to $B_0$ values of 0.42, 2.85 and 7.14\,$\mu$T). For each of the three values of the bias field $B_0$, a magnetic signal $B^x_{RF}(t)$ was applied at a specific frequency, and a spectrum analyzer was used to measure the response of the balanced polarimeter. The frequency of the $B^x_{RF}(t)$ signal was changed in 1\,kHz steps. The measurement results are shown in Fig.\,\ref{fig:Tuning}. The RF magnetometer has a full width at half maximum bandwidth of 1.45\,kHz, and a noise floor of 160\,fT/Hz$^{1/2}$, limited by the probe light noise. The polarimeter detection electronics noise (when the probe light is blocked) was measured to be below 40\,fT/Hz$^{1/2}$.

\section{\label{sec:Bloch} Polarization state of the RF magnetic field signal - Bloch vector picture}

\begin{figure}
\includegraphics[width=\columnwidth]{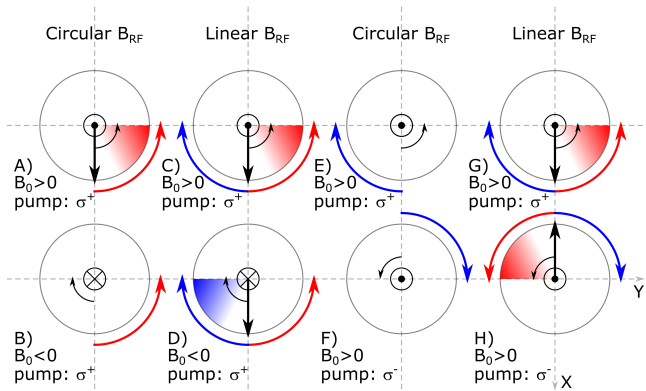}
\caption{\label{fig:Configurations} Configurations for the bias magnetic field $B_0$ direction (positive - black dot, negative - cross), polarization of the RF magnetic field $B^y_{RF}$ (text, circular or linear) and pump light circular polarization (pump: $\sigma^+$ or $\sigma^-$). The rotation direction of the RF field is shown with an large arc; the Larmor precession direction with a small arc. The initial atomic polarization component in the $xy-$plane is shown with a solid arrow, and its precession with a shaded segment.}
\end{figure}

We first analyze the response of an RF magnetometer to a resonant RF magnetic field signal of different polarization. We consider the orientation of the bias field $B_0$ that determines the magnetometer's frequency response, the direction of the circular polarization of the pump laser light, and the state of polarization of the RF magnetic field (linear or circular). In our two-channel RF magnetometer we can implement all possible configurations that are shown in Fig.\,\ref{fig:Configurations}.

The RF magnetometer's principle of operation can be intuitively described by considering the atomic polarization as a vector obeying the Bloch equations of motion. First,  consider a coordinate system with a quantization axis along the $z$-axis, and use the rotating wave approximation to transform into a system rotating with the Larmor precession frequency (determined by the value and the direction of $B_0$ as well as the atomic gyromagnetic ratio) in the corresponding Larmor precession direction \cite{AleksandrovVershovskii2009}. 

Next, assume that a $\sigma^+$ circularly-polarized laser beam creates an atomic polarization in the stretched atomic state $m_F=F$ that aligned with the $z$-axis, with $F$ the total angular momentum of the quantum system, and $m_F$ its projection along the quantization axis. A resonant rotating magnetic RF field $B_{RF}$ with a frequency equal to the Larmor precession frequency will be stationary in the coordinate system rotating in the direction of the Larmor precession, and will tilt the atomic polarization away from the quantization axis, creating an atomic polarization component in the plane orthogonal to the quantization axis ($xy$-plane). This atomic polarization will start precessing around the quantization axis. The same mechanism for driving the spin precession by an RF magnetic field is used in the classical optically-pumped M$_x$ magnetometer \cite{Bloom1962}. In our example (see Fig.\,\ref{fig:Configurations} A)), the instantaneous atomic polarization in the laboratory frame is shown in the $x$-direction with an arrow. In the stationary coordinate system, the atomic polarization component will precess at the Larmor frequency in the $xy$-plane in counter-clockwise direction. A linearly-polarized probe beam propagating in the same plane plane will experience a corresponding polarization rotation modulated at the Larmor frequency that can be detected with a polarimeter.

If the direction of the bias magnetic field $B_0$ is reversed (see Fig.\,\ref{fig:Configurations}, B)), the direction of the Larmor precession reverses. In a coordinate system rotating in the direction of the Larmor precession, the rotating RF magnetic field $\vec{B}_{RF}$ will no longer be stationary in the rotating frame, and will not cause a resonant tilt of the atomic polarization away from the quantization axis. The atomic polarization component in the $xy$-plane will be negligible compared to the case of oppositely-oriented bias field $B_0$ and no signal will be detected by the polarimeter. This situation can also be understood from the point of view of angular momentum conservation. The atoms are prepared in a stretched state with a maximum value of the projection quantum number $m_F=F$ and are not allowed to absorb a photon from the circularly-polarized RF magnetic field that would increase the value of $m_F$. 

For a linearly-polarized RF magnetic field $B_{RF}$, the situation is different (see Fig.\,\ref{fig:Configurations}, C) and D)). The linear field polarization can be decomposed into two in-phase circularly polarized field components rotating in opposite directions. For both possible orientations of the bias field $B_0$, there will be an RF magnetic field component that is stationary in the coordinate system rotating in the direction of the Larmor precession, and corresponding polarimeter signals. In both cases, a photon belonging to the corresponding circularly polarized RF field would lead to an allowed decrease of the value of $m_F$ and the corresponding signals will be out of phase.

If we construct a two-channel system, with each channel having opposite orientation of the bias magnetic field $B_0$, the sum of the polarimeter signals from the two channels will give a signal in the case of circularly-polarized RF magnetic field $B_{RF}$ (Fig.\,\ref{fig:Configurations}, A) + B)), and no signal for the case of linearly-polarized field $B_{RF}$ (Fig.\,\ref{fig:Configurations}, C) + D)). Such a noise suppression scheme relies on detecting a circularly-polarized RF magnetic signal, and rejecting a linearly-polarized noise.

Another possibility offered by the RF magnetometer is to use the direction of the circular polarization of the pump light, while keeping the direction of the bias field $B_0$ the same. If the rotation direction of a circularly-polarized RF magnetic field does not coincide with that of the Larmor precession, there is no polarimeter signal for either pump circular polarization orientation (Fig.\,\ref{fig:Configurations}, E) and F)). If the RF magnetic field is linearly polarized, the field component rotating in the direction of the Larmor precession is going to tilt the atomic polarization away from the $z$-axis. 

The atomic polarization in the $xy$-plane will either be in the direction of the $x$-axis for $\sigma^+$ circularly-polarized pump light (Fig.\,\ref{fig:Configurations}, G)) or in the opposite direction for $\sigma^-$ circularly-polarized pump light (Fig.\,\ref{fig:Configurations}, H)) at a given instant. In such a situation the difference between two channels results in either no signal (Fig.\,\ref{fig:Configurations}, E) - F)), or in twice the noise (Fig.\,\ref{fig:Configurations}, G) - H)) because there is a $\pi$\,rad phase difference for the atomic polarizations driven by a linearly-polarized noise and for $\sigma^+$ and $\sigma^-$ circularly-polarized pump beams. This channel configuration  thus leads to a measurement of only the linearly-polarized noise, rejecting the circularly-polarized RF magnetic field signal. The measured noise could then be subtracted from the output of an independent channel that measures both signal and noise.

It should be mentioned that using an RF field with the opposite circular polarization to that of the cases E) and F) of Fig.\,\ref{fig:Configurations} would result in atomic polarizations rotating out of phase with each other but in the same direction, as the bias field and Larmor precession directions are the same for both cases E) and F). This is essentially the same situation encountered in the cases G) and H). Using the difference between two such channels would result in twice the signal, but also in twice the contribution of the linearly-polarized noise.

\section{\label{sec:theory} Sensitivity to the polarization and direction of the RF magnetic field - density matrix approach}

A simple theoretical model is developed to capture qualitatively the concepts outlined in Section\,\ref{sec:Bloch} and illustrated in Fig.\,\ref{fig:Configurations}. The model considers the ground state manifold of the $^{87}$Rb atom, with the energy of the Zeeman states of the $F=2$ ground state hyperfine component determined by the Breit-Rabi formula \cite{BreitRabi1931}. The energy splitting between the Zeeman states is determined by the value of the static magnetic field $B_0$ chosen parallel to the quantization $z$-axis. The atomic system is described by a $5 \times 5$ density matrix, as all microwave transitions connecting the $F=1$ and $F=2$ components of the ground state hyperfine manifold are neglected here.

The total Hamiltonian of the system includes the atomic Hamiltonian $H_0$ in the presence of a static magnetic field $B_0$, as well as an interaction Hamiltonian $H_{RF}$ in the presence of an RF magnetic field 
\begin{align}
	\vec{B}_{RF}\left(t\right)=B^x_{RF}\left(t\right)+B^y_{RF}\left(t\right)=\\
	=B^x_{RF}\text{cos}\left(2 \pi \nu t \right)+B^y_{RF}\text{cos}\left(2 \pi \nu t + \varphi \right). \nonumber
\end{align}
confined in the $xy-$plane. The phase delay $\varphi$ between the $x-$ and $y-$components $B^x_{RF}\left(t\right)$ and $B^x_{RF}\left(t\right)$ of the RF field controls the field's polarization stage. 

The time evolution of the atomic system is calculated using the von Neumann equation:

\begin{equation}
\label{vonNeumann}
	i \hbar \frac{d \rho\left(t\right)}{ dt}=\left[\left(H_0+H_{RF}\right),\rho\left(t\right)\right].
\end{equation}
For simplicity, the atomic system evolution describing the RF magnetometer operation is broken down to three steps. 

First, the optical pumping process is taken into account by setting the proper initial conditions for the density matrix $\rho\left(0\right)$. For optical pumping with $\sigma^+$ circularly polarized light, the diagonal matrix element of $\rho\left(0\right)$ corresponding to the $\left|F=2,m_F=2\right\rangle$ atomic state is set to 1, while all others are set to zero. This corresponds to the situation when the laser light pumps all the atoms into the stretched state $m_F=F$. For the opposite circular polarization $\sigma^-$ of the optical pumping light, only the diagonal element of the initial density matrix $\rho\left(0\right)$ corresponding to $\left|F=2,m_F=-2\right\rangle$ is set to 1, while all others are set to zero.

Second, the von Neumann equation (\ref{vonNeumann}) is solved numerically. We select the following example values: $B_0=7$\,$\mu$T, $\nu=B_0\left(g_J+3g_I\right)\mu_B/\left(8 \pi \hbar \right)$, $B^x_{RF}=B^y_{RF}=9$\,nT, and the duration of the time evolution $\tau=100/\nu$. The Land\'e factor $g_J=2.002$\,$331$\,$13\left(20\right)$ for the $^{87}$Rb ground state and the nuclear $g-$factor $g_I=−0.000$\,$995$\,$141$\,$4\left(10\right)$ are taken from \cite{ARIMONDOINGUSCIOVIOLINO1977}.

Third, the final state of the atomic system $\rho\left(\tau\right)$ is used to calculate the atomic polarization along the $x-$direction
\begin{equation}
P_x=Tr\left(\rho\left(\tau\right) \hat{S}_x\right),
\end{equation}
and in the $xy-$plane,
\begin{equation}
	P_{xy}=\sqrt{\left[Tr\left(\rho\left(\tau\right) \hat{S}_x\right)\right]^2+\left[Tr\left(\rho\left(\tau\right) \hat{S}_y\right)\right]^2},
\end{equation}
with $\hat{S}_x$ and $\hat{S}_x$ the total angular projection operator in the $x-$ and $y-$directions. 

The rotation of the detection light polarization is proportional to the component of the atomic polarization along the probe beam ($x-$axis). In this way, the atomic polarization can be qualitatively compared to the output of the polarimeter, and the temporal evolution of the atomic polarization can be used to simulate the effect of the RF field on the magnetometer output. 

\begin{figure}
\includegraphics[width=\columnwidth]{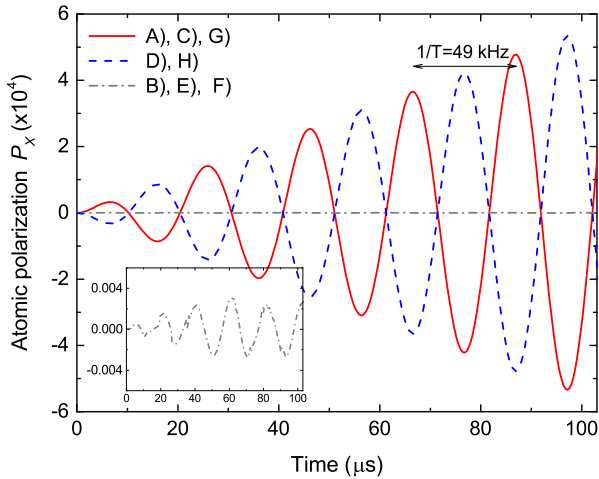}
\caption{\label{fig:Theory2} Time evolution of the atomic polarization $P_x$ for the cases shown in Fig.\,\ref{fig:Configurations}. Solid line - cases A), C) and G). Dash line - cases D) and H). Dash-dot line - cases B), E) and F). Insert: magnified vertical scale for cases B), E) and F)}
\end{figure}

Figure\,\ref{fig:Theory2} shows the amplitude of the atomic polarization $P_x$ in the $x-$direction as a function of time. The value of $P_x$ oscillates at the Larmor frequency of 49\,kHz determined by the bias field $B_0=7$\,$\mu$\,T. The RF magnetic field is chosen to have a constant magnitude, and a polarization state determined by the variable $\varphi$, according to the specific examples depicted on Fig.\,\ref{fig:Configurations}. For certain orientations of the bias field $B_0$ and the polarization state of the RF field $\vec{B}_{RF}$, the atomic polarization is negligible (cases B), E) and F)), as expected from arguments discussed in Section\,\ref{sec:Bloch}. For the cases A), C) and G), the polarization phase is determined by the direction of the counterclockwise direction of the Larmor precession for $B_0>0$ and $\sigma^+$ pump light polarization. The calculation predicts a suppression factor of $>1000$ between cases with precessing polarization and these without, as can be seen from the inset of Fig.\,\ref{fig:Theory2}. When the direction of the bias field $B_0$ is reversed, the phase of the atomic polarization precession changes by $\pi$\,rad (case D)) due to the direction change of the Larmor precession. It also precesses out of phase when $B_0>0$, when $\sigma^+$ circular polarization is used for optical pumping (case H)). 

\begin{figure}
\includegraphics[width=\columnwidth]{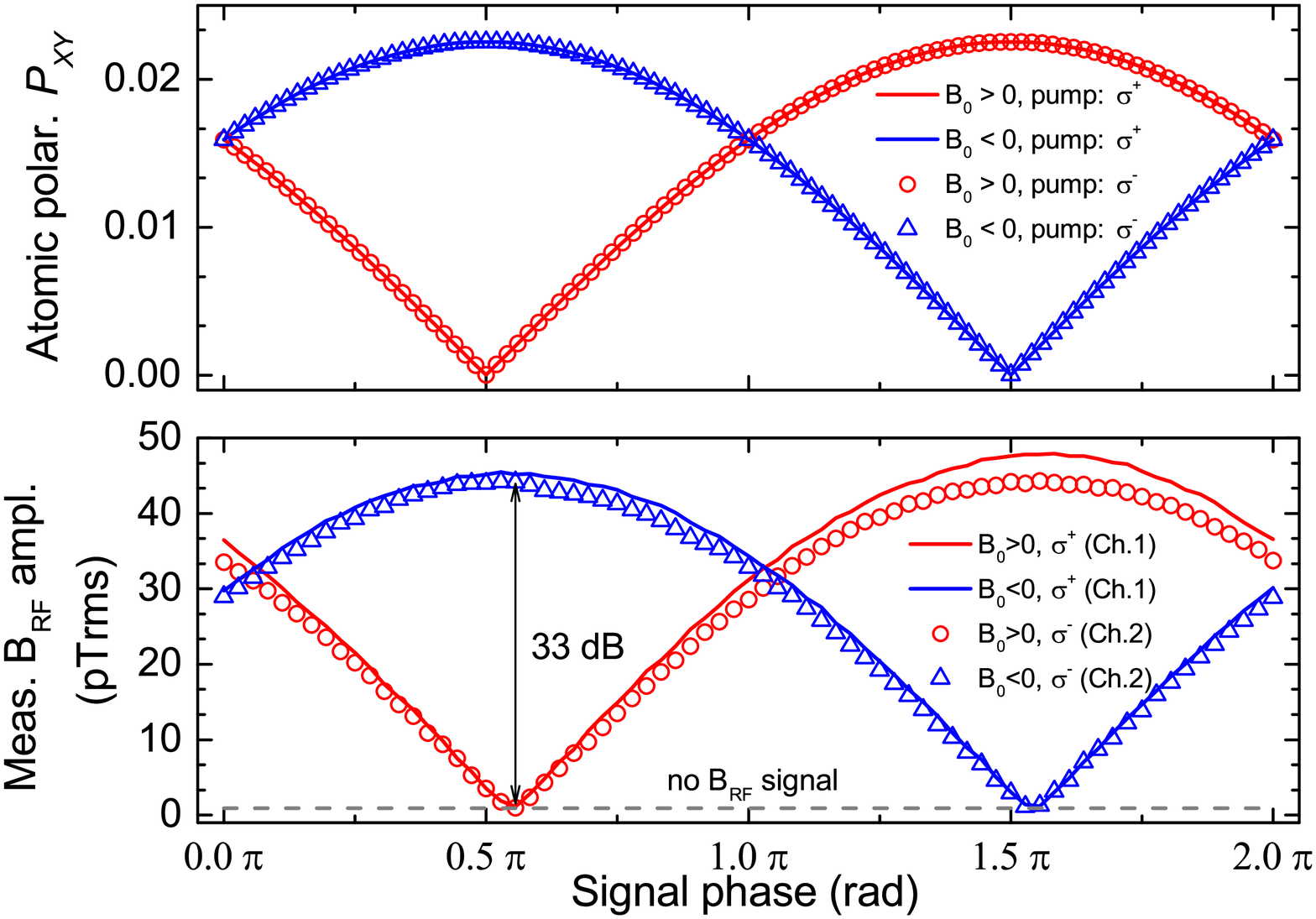}
\caption{\label{fig:Theory} Top plot: atomic polarization $P_{xy}$ as a function of the phase $\varphi$ between $B^x_{RF}\left(t\right)$ and $B^y_{RF}\left(t\right)$ after a time evolution of $\tau=100/\nu$. Red curve (red circles): $B_0>0$, pump beam polarization $\sigma^+$ ($\sigma^-$). Blue curve (blue circles): $B_0<0$, pump beam polarization $\sigma^+$ ($\sigma^-$). Bottom plot: measured polarimeter response as a function of the phase $\varphi$ between $B^x_{RF}\left(t\right)$ and $B^y_{RF}\left(t\right)$ signals at 50\,kHz. Red curve (red circles): $B_0>0$, pump beam polarization $\sigma^+$ ($\sigma^-$). Blue curve (blue circles): $B_0<0$, pump beam polarization $\sigma^+$ ($\sigma^-$).}
\end{figure}

The dependence of the atomic polarization $P_{xy}$ in the $xy-$ plane on the phase $\varphi$ controlling the RF magnetic field polarization is shown in Fig.\,\ref{fig:Theory} (top plot). Results are shown for different bias field $B_0$ directions (along or against the quantization axis direction) as well as pump light polarization ($\sigma^+$ or $\sigma^-$).

Fig.\,\ref{fig:Theory} shows that the atomic polarization $P_{xy}$ is close to zero for certain values of the phase $\varphi$. These values are $\pi/2$\,rad for orientation of $B_0$ in the positive direction of the $z-$axis, and $3 \pi/2$\,rad for $B_0$ for $B_0$ oriented along the negative direction of the $z-$axis. These values correspond to a circular RF magnetic field polarization rotating in opposite direction to that of the Larmor precession. The results verify the intuitive vector picture described in Section\,\ref{sec:Bloch} and shown in Fig.\,\ref{fig:Configurations} (cases B), E) and F)). The change of the pump light polarization between $\sigma^+$ and $\sigma^-$ does not affect the value of the $\varphi$ for which the atomic polarization $P_{xy}$ has a minimum, as the Larmor precession does not change direction. On the other hand, changing the orientation of $B_0$ changes the direction of the Larmor precession, and correspondingly to a $\pi$\,rad change of the value of $\varphi$ corresponding to the minimum value of $P_{xy}$.

Without considering the detection beam direction, the RF magnetometer has azimuthal symmetry, as the bias field $B_0$ and the optical pumping beam are parallel to the quantization axis $z$. The magnetometer is sensitive to RF magnetic fields that have a component in the $xy-$plane orthogonal to the quantization axis, and the direction of the induced atomic polarization is determined by the direction of the magnetic field vector that rotates in the direction of Larmor precession. 

\begin{figure}
\includegraphics[width=\columnwidth]{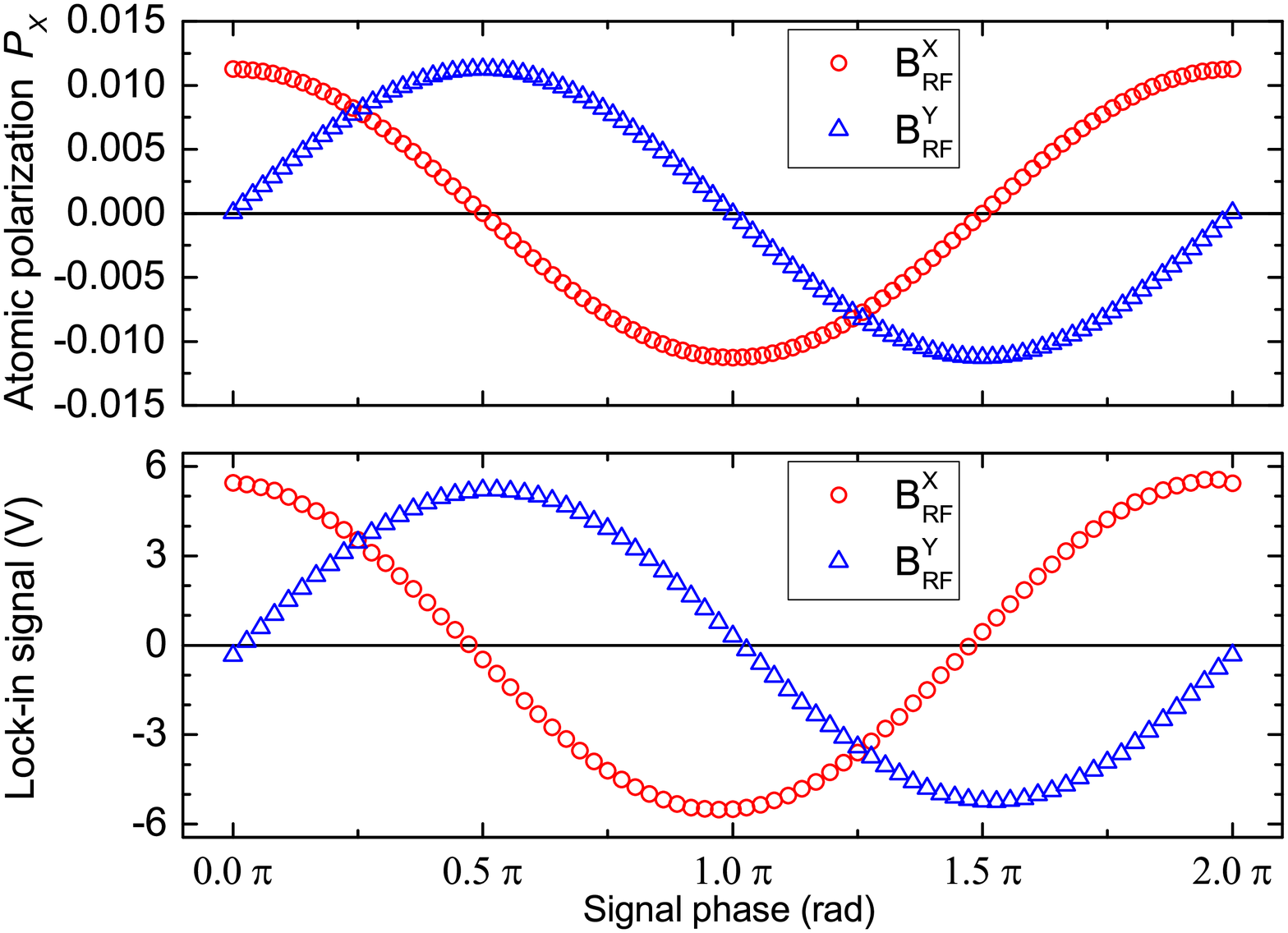}
\caption{\label{fig:Orientation} Top plot: Instantaneous amplitude of the atomic polarization $P_x$ as a function of the phase of the linearly-polarized RF field. Bottom plot: Measured in-phase signal of the detected RF field. Red circles: $B^x_{RF}$, blue triangles: $B^y_{RF}$.}
\end{figure}

As the detection beam is fixed in space, its direction with respect to the precessing atomic polarization determines the phase of the detected signal. This means that the phase difference between the RF magnetic field and the detected signal depends on the orientation of the magnetic field vector and the magnetometer detection axis at any given time. This is illustrated in Fig.\,\ref{fig:Orientation} (top plot) that shows the instantaneous amplitude of the precessing atomic polarization $P_x$ along the detection axis $x$ as a function of the phase of the linearly-polarized RF field. The RF field is applied either in the $x-$ or in the $y-$direction. The $\pi/2$\,rad phase difference acquired by $P_x$ when the RF field polarization is changed from the $x-$ to the $y-$axis demonstrates the angular phase dependence of the precessing atomic polarization $P_x$. 

\section{\label{sec:meas} Sensitivity to the polarization and direction of the RF magnetic field - measurement}

\begin{figure}
\includegraphics[width=\columnwidth]{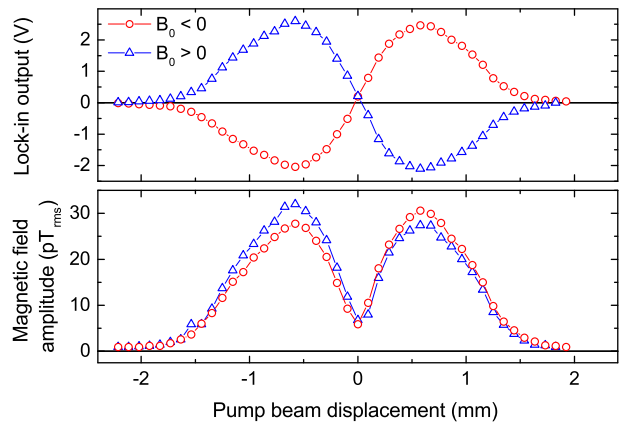}
\caption{\label{fig:Displacement} Measured phase (top plot) and amplitude (bottom plot) of the detected RF magnetic signal as a function of the pump beam displacement along the probe beam direction ($x-$axis on Fig.\,\ref{fig:Setup}). The circles (triangles) correspond to $B_0$ oriented in the positive (negative) $z-$direction.}
\end{figure}

Three sets of measurements were performed according to the intuitive picture and the density matrix calculations outlined in the previous sections. The first set studied the effect of the pump light polarization on the magnetometer response. The pump laser beam was collimated to a diameter less than the width of a single magnetometer channel of 1.5\,mm. Using a mirror mounted on a precision rotation mount and positioned $\sim$60\,cm away from the magnetometer, the direction of the pump beam was changed by a small angle. The angle change resulted in a calibrated spatial displacement along the direction of the probe beam, thus creating spatially-varying optical pumping as the pump beam illuminated one or the other magnetometer channels. The magnetometer was tuned to 50\,kHz, and a 50\,kHz signal was applied along the $x-$direction. The 50\,kHz signal at the polarimeter output was detected with a lock-in amplifier and a spectrum analyzer, providing phase and amplitude information of the detected 50\,kHz signal. The signal amplitude was converted in magnetic field values by measuring the ac voltage drop across a resistor connected in series with the field coils. The detected RF signal phase and amplitude as a function of pump beam displacement are shown in Fig.\,\ref{fig:Displacement}.

The results show that as the pump beam is scanned across each of the magnetometer channels, the maximum amplitude of the detected RF signal is similar for both channels, regardless of the orientation of the bias field $B_0$. At the same time, each channel detects opposite signal phase, as expected from the cases G) and H) (see Fig.\,\ref{fig:Configurations} as well as Fig.\,\ref{fig:Theory2}). The detected signal amplitude corresponds to the amplitude of the precessing atomic polarization $P_{xy}$ in the magnetometer's $xy-$plane, while the phase is related to the atomic polarization $P_x$ along the magnetometer's probe beam direction ($x-$axis). The small differences in the response of the individual channels to the pump beam displacement or the magnetic field orientation is probably due to slight differences in the magnetometer channels, imperfect alignment of the direction and polarization of the bias and RF magnetic fields, as well as the direction and polarization of the pump and probe laser beams. 

The measurements show that the precessing atomic polarization $P_{xy}$ has an opposite orientation for opposite circular polarizations of the pump light. This sensitivity can be used to sum the output of two magnetometer channels that would result in a measurement of a linearly polarized RF noise (or signal). 

The second set of measurements studied the effect of the RF field polarization on the magnetometer response. The pump laser beam had a diameter less than the width of a single magnetometer channel of 1.5\,mm, and was optically pumping either the first ($\sigma^+$) or the second ($\sigma^-$) magnetometer channel. Two RF magnetic signals $B^x_{RF}$ and $B^y_{RF}$ of the same frequency were applied simultaneously. The two 50\,kHz signals were derived from a two-channel function generator. The phase $\varphi$ of the $B^y_{RF}$ signal was changed from $0$ to $2 \pi$\,rad in $5^{\circ}$ (87\,mrad) steps. A spectrum analyzer connected to the polarimeter output was used to measure the amplitude of the polarimeter response at 50\,kHz that is proportional to the magnitude of the precessing atomic polarization $P_{xy}$ in the $xy-$plane. The measurements are shown in Fig.\,\ref{fig:Theory} (bottom plot). They show a minimum amplitude of the detected RF signal for values of $\varphi$ in the vicinity of $\pi/2$\,rad and $3 \pi/2$\,rad. The ratio of the maximum to minimum values close to $\varphi=\pi/2$\,rad is $44$, corresponding to 33\,dB rejection of the corresponding circular RF field polarization. The minimum value is comparable to the measurement noise floor (without an applied RF signal), and it is expected that the rejection would be even higher. It is also expected that the rejection depends on the difference between the Larmor precession and the magnetic signal frequencies.

The rotation of the probe beam polarization is proportional to the atomic polarization along the direction of the probe beam. It it beyond the scope of this work to compare quantitatively the calculation and measurement results shown in Fig.\,\ref{fig:Theory}; the results nevertheless demonstrate that for certain values of the phase $\varphi$ corresponding to circularly-polarized RF magnetic field, the RF magnetometer has a minimum or a maximum sensitivity to the RF field depending on the direction of the bias field $B_0$. This sensitivity to the polarization of the RF field can be explored to reject RF magnetic noise or signals of linear polarization while keeping the circularly-polarized RF signal unaffected by forming the difference between two magnetometer channels with opposite direction of the bias field $B_0$.

\begin{figure}
\includegraphics[width=\columnwidth]{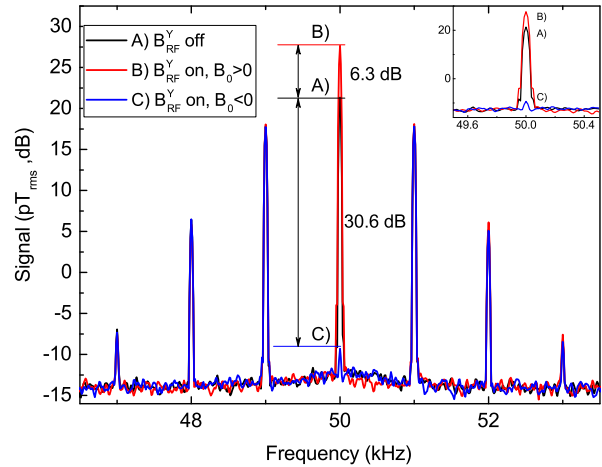}
\caption{\label{fig:PM} Measured signal amplitude as a function of the signal frequency. Case A): PM-modulated $B^x_{RF}$ signal only. Case B): PM-modulated $B^x_{RF}$ signal and in-phase reference signal $B^y_{RF}$. Case C): PM-modulated $B^x_{RF}$ signal and quadrature reference signal $B^y_{RF}$. The insert shows the 50\,kHz carrier with reduced frequency span.}
\end{figure}

The measured difference in the sensor's response to a linearly- and circularly-polarized RF magnetic field is demonstrated in Figure\,\ref{fig:PM}. As a signal $B^x_{RF}$, we use a phase-modulated (PM) tone at 50\,kHz, with 1\,kHz modulation frequency and a deviation $\pi/2$\,rad. As a reference $B^y_{RF}$, we use a single tone at 50\,kHz. With the reference signal turned of, the black curve A) shows the typical spectrum consisting of a carrier at 50\,kHz and sidebands separated by 1\,kHz with amplitude dependent on the modulation index and the magnetometer's 1.5\,kHz frequency bandwidth. When the reference signal $B^y_{RF}$ is turned on, for a certain value of its phase the amplitude of the 50\,kHz carrier is increased by $\sim$6\,dB as shown by the red curve B) - the 50\,kHz signal becomes circularly polarized and rotates in the same direction as the Larmor precession. When the direction of the bias magnetic field $B_0$ is reversed, the 50\,kHz signal is suppressed by $>30$\,dB as shown by the blue curve C) - the signal is again circularly-polarized but rotates in opposite direction. The PM sidebands remain linearly-polarized in all three cases, and are largely unaffected by the bias field direction switching. The measured difference in the detected signal amplitudes for opposite circularly-polarized RF signals is $>36$\,dB, showing that linearly-polarized noise can be suppressed by $>30$\,dB. 

The results from the third set of measurements demonstrate the azimuthal sensitivity of the RF magnetometer. A linearly-polarized RF signal at 50\,kHz from the function generator was applied either in the $x-$ or in the $y-$direction only. The amplitude of the detected signal at the polarimeter output was measured as a function of the phase of the applied signal using a lock-in amplifier referenced by the function generator. The result is shown in Fig.\,\ref{fig:Orientation} (bottom plot). The sinusoidal dependence of the signal amplitude shows the linear relationship between the phases of the applied and detected signals. The $\pi/2$\,rad phase difference in the detected signal for RF field applied along the $x-$ or $y-$directions shows the sensitivity of the RF magnetometer to the detection axis orientation with respect to the polarization direction of the RF field signal. 

\section{Discussion}

Based on the results from Sections \ref{sec:Bloch} and \ref{sec:meas}, the proposed balanced schemes can be used to construct a low-frequency magnetic field sensor designed to detect circularly-polarized signals while suppressing linearly-polarized ambient noise. The tunability of the RF magnetometer with the value of the bias magnetic field $B_0$ can be used to create a sensor operating at a specific frequency in the range of $\sim$1\,kHz up to 1\,MHz and a bandwidth in the range of 1\,kHz. The sensor can be used to detect circularly-polarized magnetic fields and reject linearly-polarized ambient noise by either the use of a channel pair having opposite direction of the bias field $B_0$, or with $\sigma^+$ and $\sigma^-$ circular polarization of the optical pumping light. The necessary difference or sum of the channel pair outputs can be done electronically, but also optically. The sign reversal can be accomplished by an additional $\pi/2$\,rad rotation of a channel's probe light polarization, after the light-atom interaction but before the light enters the channel's polarimeter using a half-waveplate. In the case of bias field reversal, the necessary sum of the channel pair outputs can be done optically by allowing the probe light to propagate consecutively through two spatial vapor cell regions with opposite directions of $B_0$. 

The discussed RF magnetometer features depend on the polarization state of the RF signals. In practice, the detected RF signals are typically not circularly-polarized. One solution would be to use multiple transmitters to create such signals, which might not be practical. The requirement for circular signal polarization limits the application of the noise suppression techniques discussed in this work to the case of artificially created signals, such as ones used in magnetic field communications \cite{GerginovSilvaHowe2017}. It might be possible to apply the technique to free induction decay signals of specific frequency excited by an electromagnetic pulse (such as ultra-low field NMR \cite{SavukovSeltzerRomalis2007} and Nuclear Quadrupole Resonance \cite{LeeSauerSeltzerEtAl2006}). Another possible application is in magnetic induction imaging, where the presence of conductive objects changes the phase of the detected magnetic signal \cite{DeansMarmugiHussainEtAl2016}. Finally, as the optically-pumped medium absorbs only one of the two posible circularly-polarized RF field components, one can use this feature to construct an atomic circularly polarized antenna.

As shown in this work, one can change the polarization properties of an effective RF signal by combining a linearly-polarized remote signal with a local reference RF signal of the same frequency and with a specific amplitude and phase relation between remote and reference signals. For example, the signal $B^x_{RF}\left(t\right)$ used in this work can be treated as coming from a remote source, and the signal $B^y_{RF}\left(t\right)$ as reference. If the frequency of the remote signal $B^x_{RF}\left(t\right)$ is known, by adjusting the amplitude and phase of the reference signal $B^y_{RF}\left(t\right)$ one can achieve the desired maximum sensitivity difference shown in Fig.\,\ref{fig:Theory}. The experimentally demonstrated amplitude rejection factor of 63 (36\,dB) of the circularly-polarized RF signal rotating against the direction of Larmor precession would allow a factor of 30 ambient noise suppression. For communications, an order of magnitude lower noise would either reduce the transmitted RF signal power by a factor of 100, or increase the communications range by a factor of two (for magnetic-dipole type signal attenuation with range), or decrease the averaging time by a factor of 100 \cite{GerginovSilvaHowe2017}.

The demonstrated azimuthal dependence of the detected signal phase can be used to gain information about the magnetometer's orientation with respect to the source of an RF signal. As has been demonstrated already, the signal phase acquired with an atomic magnetometer could reach 1\,mrad level with averaging \cite{GerginovSilvaHowe2017}. It is expected that a similar level of precision could be achieved with regard to the azimuthal angle using an RF magnetometer.

\section{Conclusions}

It is shown both theoretically and experimentally that an optically pumped RF atomic magnetometer has a minimum or a maximum sensitivity to the direction of a circularly-polarized RF signal, depending on the direction of the Larmor precession determined by the magnetometer's bias field. The sensitivity can be tuned by reversing the direction of the magnetometer's bias field. Depending on the direction of the circular polarization of the optically pumping light, the atomic polarization precession acquires a $\pi$\,rad phase change. These results allow one to realize balanced detection schemes that measure circularly-polarized RF signals and suppress linearly-polarized noise, or measure the linearly-polarized noise and suppress circularly-polarized signals. These noise suppression schemes do not depend on the relative distance between the signal source, the noise source, and the sensor, and can be realized electronically or optically.

\section{Acknowledgements}
 
The authors thank Fabio da Silva, Craig Nelson, Archita Hati, David Howe and Susan Schima for the helpful discussions and technical help, and Jeff Sherman and Tom Heavner for their helpful comments on the manuscript. This work is a contribution of NIST, an agency of the US government, and is not subject to copyright in the US. 

\section{References}


\begin{thebibliography}{21}%
\makeatletter
\providecommand \@ifxundefined [1]{%
 \@ifx{#1\undefined}
}%
\providecommand \@ifnum [1]{%
 \ifnum #1\expandafter \@firstoftwo
 \else \expandafter \@secondoftwo
 \fi
}%
\providecommand \@ifx [1]{%
 \ifx #1\expandafter \@firstoftwo
 \else \expandafter \@secondoftwo
 \fi
}%
\providecommand \natexlab [1]{#1}%
\providecommand \enquote  [1]{``#1''}%
\providecommand \bibnamefont  [1]{#1}%
\providecommand \bibfnamefont [1]{#1}%
\providecommand \citenamefont [1]{#1}%
\providecommand \href@noop [0]{\@secondoftwo}%
\providecommand \href [0]{\begingroup \@sanitize@url \@href}%
\providecommand \@href[1]{\@@startlink{#1}\@@href}%
\providecommand \@@href[1]{\endgroup#1\@@endlink}%
\providecommand \@sanitize@url [0]{\catcode `\\12\catcode `\$12\catcode
  `\&12\catcode `\#12\catcode `\^12\catcode `\_12\catcode `\%12\relax}%
\providecommand \@@startlink[1]{}%
\providecommand \@@endlink[0]{}%
\providecommand \url  [0]{\begingroup\@sanitize@url \@url }%
\providecommand \@url [1]{\endgroup\@href {#1}{\urlprefix }}%
\providecommand \urlprefix  [0]{URL }%
\providecommand \Eprint [0]{\href }%
\providecommand \doibase [0]{http://dx.doi.org/}%
\providecommand \selectlanguage [0]{\@gobble}%
\providecommand \bibinfo  [0]{\@secondoftwo}%
\providecommand \bibfield  [0]{\@secondoftwo}%
\providecommand \translation [1]{[#1]}%
\providecommand \BibitemOpen [0]{}%
\providecommand \bibitemStop [0]{}%
\providecommand \bibitemNoStop [0]{.\EOS\space}%
\providecommand \EOS [0]{\spacefactor3000\relax}%
\providecommand \BibitemShut  [1]{\csname bibitem#1\endcsname}%
\let\auto@bib@innerbib\@empty
\bibitem [{\citenamefont {Sheng}\ \emph {et~al.}(2013)\citenamefont {Sheng},
  \citenamefont {Li}, \citenamefont {Dural},\ and\ \citenamefont
  {Romalis}}]{ShengLiDuralEtAl2013}%
  \BibitemOpen
  \bibfield  {author} {\bibinfo {author} {\bibfnamefont {D.}~\bibnamefont
  {Sheng}}, \bibinfo {author} {\bibfnamefont {S.}~\bibnamefont {Li}}, \bibinfo
  {author} {\bibfnamefont {N.}~\bibnamefont {Dural}}, \ and\ \bibinfo {author}
  {\bibfnamefont {M.~V.}\ \bibnamefont {Romalis}},\ }\bibfield  {title}
  {\enquote {\bibinfo {title} {Subfemtotesla scalar atomic magnetometry using
  multipass cells},}\ }\href {\doibase 10.1103/PhysRevLett.110.160802}
  {\bibfield  {journal} {\bibinfo  {journal} {Phys. Rev. Lett.}\ }\textbf
  {\bibinfo {volume} {110}},\ \bibinfo {pages} {160802} (\bibinfo {year}
  {2013})}\BibitemShut {NoStop}%
\bibitem [{\citenamefont {Lee}\ \emph {et~al.}(2006)\citenamefont {Lee},
  \citenamefont {Sauer}, \citenamefont {Seltzer}, \citenamefont {Alem},\ and\
  \citenamefont {Romalis}}]{LeeSauerSeltzerEtAl2006}%
  \BibitemOpen
  \bibfield  {author} {\bibinfo {author} {\bibfnamefont {S.-K.}\ \bibnamefont
  {Lee}}, \bibinfo {author} {\bibfnamefont {K.~L.}\ \bibnamefont {Sauer}},
  \bibinfo {author} {\bibfnamefont {S.~J.}\ \bibnamefont {Seltzer}}, \bibinfo
  {author} {\bibfnamefont {O.}~\bibnamefont {Alem}}, \ and\ \bibinfo {author}
  {\bibfnamefont {M.~V.}\ \bibnamefont {Romalis}},\ }\bibfield  {title}
  {\enquote {\bibinfo {title} {Subfemtotesla radio-frequency atomic
  magnetometer for detection of nuclear quadrupole resonance},}\ }\href@noop {}
  {\bibfield  {journal} {\bibinfo  {journal} {Applied Physics Letters}\
  }\textbf {\bibinfo {volume} {89}},\ \bibinfo {pages} {214106} (\bibinfo
  {year} {2006})}\BibitemShut {NoStop}%
\bibitem [{\citenamefont {Clarke}(2010)}]{Clarke2010}%
  \BibitemOpen
  \bibfield  {author} {\bibinfo {author} {\bibfnamefont {John}\ \bibnamefont
  {Clarke}},\ }\bibfield  {title} {\enquote {\bibinfo {title} {{SQUIDs: Then
  and now}},}\ }\href
  {http://search.ebscohost.com.proxy.library.nd.edu/login.aspx?direct=true&db=aph&AN=54478592&site=ehost-live}
  {\bibfield  {journal} {\bibinfo  {journal} {International Journal of Modern
  Physics B: Condensed Matter Physics; Statistical Physics; Applied Physics}\
  }\textbf {\bibinfo {volume} {24}},\ \bibinfo {pages} {3999 -- 4038} (\bibinfo
  {year} {2010})}\BibitemShut {NoStop}%
\bibitem [{\citenamefont {Seltzer}\ and\ \citenamefont
  {Romalis}(2004)}]{SeltzerRomalis2004}%
  \BibitemOpen
  \bibfield  {author} {\bibinfo {author} {\bibfnamefont {S.~J.}\ \bibnamefont
  {Seltzer}}\ and\ \bibinfo {author} {\bibfnamefont {M.~V.}\ \bibnamefont
  {Romalis}},\ }\bibfield  {title} {\enquote {\bibinfo {title} {Unshielded
  three-axis vector operation of a spin-exchange-relaxation-free atomic
  magnetometer},}\ }\href {\doibase http://dx.doi.org/10.1063/1.1814434}
  {\bibfield  {journal} {\bibinfo  {journal} {Applied Physics Letters}\
  }\textbf {\bibinfo {volume} {85}},\ \bibinfo {pages} {4804--4806} (\bibinfo
  {year} {2004})}\BibitemShut {NoStop}%
\bibitem [{\citenamefont {Slack}\ \emph {et~al.}(1967)\citenamefont {Slack},
  \citenamefont {Vance},\ and\ \citenamefont {Lynch}}]{SlackVanceLynch1967}%
  \BibitemOpen
  \bibfield  {author} {\bibinfo {author} {\bibfnamefont {H.}~\bibnamefont
  {Slack}}, \bibinfo {author} {\bibfnamefont {M.}~\bibnamefont {Vance}}, \ and\
  \bibinfo {author} {\bibfnamefont {L.~L.}\ \bibnamefont {Lynch}},\ }\bibfield
  {title} {\enquote {\bibinfo {title} {The geomagnetic gradiometer},}\
  }\href@noop {} {\bibfield  {journal} {\bibinfo  {journal} {Geophysics}\
  }\textbf {\bibinfo {volume} {32}},\ \bibinfo {pages} {877--892} (\bibinfo
  {year} {1967})}\BibitemShut {NoStop}%
\bibitem [{\citenamefont {Constable}(2016)}]{Constable2016}%
  \BibitemOpen
  \bibfield  {author} {\bibinfo {author} {\bibfnamefont {Catherine}\
  \bibnamefont {Constable}},\ }\bibfield  {title} {\enquote {\bibinfo {title}
  {Earth's electromagnetic environment},}\ }\href {\doibase
  10.1007/s10712-015-9351-1} {\bibfield  {journal} {\bibinfo  {journal}
  {Surveys in Geophysics}\ }\textbf {\bibinfo {volume} {37}},\ \bibinfo {pages}
  {27--45} (\bibinfo {year} {2016})}\BibitemShut {NoStop}%
\bibitem [{\citenamefont {Cooper}\ \emph {et~al.}(2018)\citenamefont {Cooper},
  \citenamefont {Prescott}, \citenamefont {Lee},\ and\ \citenamefont
  {Sauer}}]{CooperPrescottLeeEtAl2018}%
  \BibitemOpen
  \bibfield  {author} {\bibinfo {author} {\bibfnamefont {Robert~J.}\
  \bibnamefont {Cooper}}, \bibinfo {author} {\bibfnamefont {David~W.}\
  \bibnamefont {Prescott}}, \bibinfo {author} {\bibfnamefont {Garrett~J.}\
  \bibnamefont {Lee}}, \ and\ \bibinfo {author} {\bibfnamefont {Karen~L.}\
  \bibnamefont {Sauer}},\ }\bibfield  {title} {\enquote {\bibinfo {title} {{RF
  atomic magnetometer array with over 40\,dB interference suppression using
  electron spin resonance}},}\ }\href {\doibase
  https://doi.org/10.1016/j.jmr.2018.08.007} {\bibfield  {journal} {\bibinfo
  {journal} {Journal of Magnetic Resonance}\ }\textbf {\bibinfo {volume}
  {296}},\ \bibinfo {pages} {36 -- 46} (\bibinfo {year} {2018})}\BibitemShut
  {NoStop}%
\bibitem [{\citenamefont {Slocum}\ and\ \citenamefont
  {Reilly}(1963)}]{SlocumReilly1963}%
  \BibitemOpen
  \bibfield  {author} {\bibinfo {author} {\bibfnamefont {R.~E.}\ \bibnamefont
  {Slocum}}\ and\ \bibinfo {author} {\bibfnamefont {F.~N.}\ \bibnamefont
  {Reilly}},\ }\bibfield  {title} {\enquote {\bibinfo {title} {Low field helium
  magnetometer for space applications},}\ }\href {\doibase
  10.1109/TNS.1963.4323257} {\bibfield  {journal} {\bibinfo  {journal} {IEEE
  Transactions on Nuclear Science}\ }\textbf {\bibinfo {volume} {10}},\
  \bibinfo {pages} {165--171} (\bibinfo {year} {1963})}\BibitemShut {NoStop}%
\bibitem [{\citenamefont {Livanov}\ \emph {et~al.}(1981)\citenamefont
  {Livanov}, \citenamefont {Koslov}, \citenamefont {Sinelnikova}, \citenamefont
  {Kholodov}, \citenamefont {Markin}, \citenamefont {Gorbach},\ and\
  \citenamefont {Korinewsky}}]{LivanovKoslovSinelnikovaEtAl1981}%
  \BibitemOpen
  \bibfield  {author} {\bibinfo {author} {\bibfnamefont {M.~N.}\ \bibnamefont
  {Livanov}}, \bibinfo {author} {\bibfnamefont {A.~N.}\ \bibnamefont {Koslov}},
  \bibinfo {author} {\bibfnamefont {S.~E.}\ \bibnamefont {Sinelnikova}},
  \bibinfo {author} {\bibfnamefont {J.~A.}\ \bibnamefont {Kholodov}}, \bibinfo
  {author} {\bibfnamefont {V.~P.}\ \bibnamefont {Markin}}, \bibinfo {author}
  {\bibfnamefont {A.~M.}\ \bibnamefont {Gorbach}}, \ and\ \bibinfo {author}
  {\bibfnamefont {A.~V.}\ \bibnamefont {Korinewsky}},\ }\bibfield  {title}
  {\enquote {\bibinfo {title} {Record of the human magnetocardiogram by the
  quantum gradiometer with optical pumping},}\ }\href@noop {} {\bibfield
  {journal} {\bibinfo  {journal} {Advances in cardiology}\ }\textbf {\bibinfo
  {volume} {28}} (\bibinfo {year} {1981})}\BibitemShut {NoStop}%
\bibitem [{\citenamefont {Savukov}\ \emph {et~al.}(2005)\citenamefont
  {Savukov}, \citenamefont {Seltzer}, \citenamefont {Romalis},\ and\
  \citenamefont {Sauer}}]{SavukovSeltzerRomalisEtAl2005}%
  \BibitemOpen
  \bibfield  {author} {\bibinfo {author} {\bibfnamefont {I.~M.}\ \bibnamefont
  {Savukov}}, \bibinfo {author} {\bibfnamefont {S.~J.}\ \bibnamefont
  {Seltzer}}, \bibinfo {author} {\bibfnamefont {M.~V.}\ \bibnamefont
  {Romalis}}, \ and\ \bibinfo {author} {\bibfnamefont {K.~L.}\ \bibnamefont
  {Sauer}},\ }\bibfield  {title} {\enquote {\bibinfo {title} {Tunable atomic
  magnetometer for detection of radio-frequency magnetic fields},}\ }\href
  {\doibase 10.1103/PhysRevLett.95.063004} {\bibfield  {journal} {\bibinfo
  {journal} {Phys. Rev. Lett.}\ }\textbf {\bibinfo {volume} {95}},\ \bibinfo
  {pages} {063004} (\bibinfo {year} {2005})}\BibitemShut {NoStop}%
\bibitem [{\citenamefont {Oida}\ \emph {et~al.}(2012)\citenamefont {Oida},
  \citenamefont {Ito}, \citenamefont {Kamada},\ and\ \citenamefont
  {Kobayashi}}]{OidaItoKamadaEtAl2012}%
  \BibitemOpen
  \bibfield  {author} {\bibinfo {author} {\bibfnamefont {Takenori}\
  \bibnamefont {Oida}}, \bibinfo {author} {\bibfnamefont {Yosuke}\ \bibnamefont
  {Ito}}, \bibinfo {author} {\bibfnamefont {Keigo}\ \bibnamefont {Kamada}}, \
  and\ \bibinfo {author} {\bibfnamefont {Tetsuo}\ \bibnamefont {Kobayashi}},\
  }\bibfield  {title} {\enquote {\bibinfo {title} {Detecting rotating magnetic
  fields using optically pumped atomic magnetometers for measuring
  ultra-low-field magnetic resonance signals},}\ }\href {\doibase
  https://doi.org/10.1016/j.jmr.2012.01.015} {\bibfield  {journal} {\bibinfo
  {journal} {Journal of Magnetic Resonance}\ }\textbf {\bibinfo {volume}
  {217}},\ \bibinfo {pages} {6 -- 9} (\bibinfo {year} {2012})}\BibitemShut
  {NoStop}%
\bibitem [{\citenamefont {Gerginov}\ \emph {et~al.}(2017)\citenamefont
  {Gerginov}, \citenamefont {da~Silva},\ and\ \citenamefont
  {Howe}}]{GerginovSilvaHowe2017}%
  \BibitemOpen
  \bibfield  {author} {\bibinfo {author} {\bibfnamefont {V.}~\bibnamefont
  {Gerginov}}, \bibinfo {author} {\bibfnamefont {F.~C.~S.}\ \bibnamefont
  {da~Silva}}, \ and\ \bibinfo {author} {\bibfnamefont {D.}~\bibnamefont
  {Howe}},\ }\bibfield  {title} {\enquote {\bibinfo {title} {Prospects for
  magnetic field communications and location using quantum sensors},}\
  }\href@noop {} {\bibfield  {journal} {\bibinfo  {journal} {Review of
  Scientific Instruments}\ }\textbf {\bibinfo {volume} {88}},\ \bibinfo {pages}
  {125005} (\bibinfo {year} {2017})}\BibitemShut {NoStop}%
\bibitem [{\citenamefont {Savukov}\ and\ \citenamefont
  {Romalis}(2005)}]{SavukovRomalis2005}%
  \BibitemOpen
  \bibfield  {author} {\bibinfo {author} {\bibfnamefont {I.~M.}\ \bibnamefont
  {Savukov}}\ and\ \bibinfo {author} {\bibfnamefont {M.~V.}\ \bibnamefont
  {Romalis}},\ }\bibfield  {title} {\enquote {\bibinfo {title} {Nmr detection
  with an atomic magnetometer},}\ }\href {\doibase
  10.1103/PhysRevLett.94.123001} {\bibfield  {journal} {\bibinfo  {journal}
  {Phys. Rev. Lett.}\ }\textbf {\bibinfo {volume} {94}},\ \bibinfo {pages}
  {123001} (\bibinfo {year} {2005})}\BibitemShut {NoStop}%
\bibitem [{\citenamefont {Deans}\ \emph {et~al.}(2016)\citenamefont {Deans},
  \citenamefont {Marmugi}, \citenamefont {Hussain},\ and\ \citenamefont
  {Renzoni}}]{DeansMarmugiHussainEtAl2016}%
  \BibitemOpen
  \bibfield  {author} {\bibinfo {author} {\bibfnamefont {Cameron}\ \bibnamefont
  {Deans}}, \bibinfo {author} {\bibfnamefont {Luca}\ \bibnamefont {Marmugi}},
  \bibinfo {author} {\bibfnamefont {Sarah}\ \bibnamefont {Hussain}}, \ and\
  \bibinfo {author} {\bibfnamefont {Ferruccio}\ \bibnamefont {Renzoni}},\
  }\bibfield  {title} {\enquote {\bibinfo {title} {Electromagnetic induction
  imaging with a radio-frequency atomic magnetometer},}\ }\href {\doibase
  10.1063/1.4943659} {\bibfield  {journal} {\bibinfo  {journal} {Applied
  Physics Letters}\ }\textbf {\bibinfo {volume} {108}},\ \bibinfo {pages}
  {103503} (\bibinfo {year} {2016})}\BibitemShut {NoStop}%
\bibitem [{dis()}]{disclaimer}%
  \BibitemOpen
  \href@noop {} {}\bibinfo {howpublished} {Any mention of commercial products
  is for information only; it does not imply recommendation or endorsement by
  NIST.}\BibitemShut {Stop}%
\bibitem [{\citenamefont {Budker}\ and\ \citenamefont
  {Romalis}(2007)}]{BudkerRomalis2007}%
  \BibitemOpen
  \bibfield  {author} {\bibinfo {author} {\bibfnamefont {Dmitry}\ \bibnamefont
  {Budker}}\ and\ \bibinfo {author} {\bibfnamefont {Michael}\ \bibnamefont
  {Romalis}},\ }\bibfield  {title} {\enquote {\bibinfo {title} {Optical
  magnetometry},}\ }\href {http://dx.doi.org/10.1038/nphys566} {\bibfield
  {journal} {\bibinfo  {journal} {Nat Phys}\ }\textbf {\bibinfo {volume} {3}},\
  \bibinfo {pages} {227--234} (\bibinfo {year} {2007})}\BibitemShut {NoStop}%
\bibitem [{\citenamefont {Aleksandrov}\ and\ \citenamefont
  {Vershovskii}(2009)}]{AleksandrovVershovskii2009}%
  \BibitemOpen
  \bibfield  {author} {\bibinfo {author} {\bibfnamefont {E.~B.}\ \bibnamefont
  {Aleksandrov}}\ and\ \bibinfo {author} {\bibfnamefont {A.~K.}\ \bibnamefont
  {Vershovskii}},\ }\bibfield  {title} {\enquote {\bibinfo {title} {Modern
  radio-optical methods in quantum magnetometry},}\ }\href@noop {} {\bibfield
  {journal} {\bibinfo  {journal} {Physics-Uspekhi}\ }\textbf {\bibinfo {volume}
  {52}},\ \bibinfo {pages} {573--601} (\bibinfo {year} {2009})}\BibitemShut
  {NoStop}%
\bibitem [{\citenamefont {Bloom}(1962)}]{Bloom1962}%
  \BibitemOpen
  \bibfield  {author} {\bibinfo {author} {\bibfnamefont {Arnold~L.}\
  \bibnamefont {Bloom}},\ }\bibfield  {title} {\enquote {\bibinfo {title}
  {Principles of operation of the rubidium vapor magnetometer},}\ }\href
  {\doibase 10.1364/AO.1.000061} {\bibfield  {journal} {\bibinfo  {journal}
  {Appl. Opt.}\ }\textbf {\bibinfo {volume} {1}},\ \bibinfo {pages} {61--68}
  (\bibinfo {year} {1962})}\BibitemShut {NoStop}%
\bibitem [{\citenamefont {Breit}\ and\ \citenamefont
  {Rabi}(1931)}]{BreitRabi1931}%
  \BibitemOpen
  \bibfield  {author} {\bibinfo {author} {\bibfnamefont {G}~\bibnamefont
  {Breit}}\ and\ \bibinfo {author} {\bibfnamefont {II}~\bibnamefont {Rabi}},\
  }\bibfield  {title} {\enquote {\bibinfo {title} {Measurement of nuclear
  spin},}\ }\href {\doibase 10.1103/PhysRev.38.2082.2} {\bibfield  {journal}
  {\bibinfo  {journal} {Phys. Rev.}\ }\textbf {\bibinfo {volume} {38}},\
  \bibinfo {pages} {2082--2083} (\bibinfo {year} {1931})}\BibitemShut {NoStop}%
\bibitem [{\citenamefont {Arimondo}\ \emph {et~al.}(1977)\citenamefont
  {Arimondo}, \citenamefont {Inguscio},\ and\ \citenamefont
  {Violino}}]{ARIMONDOINGUSCIOVIOLINO1977}%
  \BibitemOpen
  \bibfield  {author} {\bibinfo {author} {\bibfnamefont {E.}~\bibnamefont
  {Arimondo}}, \bibinfo {author} {\bibfnamefont {M.}~\bibnamefont {Inguscio}},
  \ and\ \bibinfo {author} {\bibfnamefont {P.}~\bibnamefont {Violino}},\
  }\bibfield  {title} {\enquote {\bibinfo {title} {Experimental determinations
  of hyperfine-structure in alkali atoms},}\ }\href {\doibase
  {10.1103/RevModPhys.49.31}} {\bibfield  {journal} {\bibinfo  {journal} {Rev.
  Mod. Phys.}\ }\textbf {\bibinfo {volume} {49}},\ \bibinfo {pages} {31--75}
  (\bibinfo {year} {1977})}\BibitemShut {NoStop}%
\bibitem [{\citenamefont {Savukov}\ \emph {et~al.}(2007)\citenamefont
  {Savukov}, \citenamefont {Seltzer},\ and\ \citenamefont
  {Romalis}}]{SavukovSeltzerRomalis2007}%
  \BibitemOpen
  \bibfield  {author} {\bibinfo {author} {\bibfnamefont {I.M.}\ \bibnamefont
  {Savukov}}, \bibinfo {author} {\bibfnamefont {S.J.}\ \bibnamefont {Seltzer}},
  \ and\ \bibinfo {author} {\bibfnamefont {M.V.}\ \bibnamefont {Romalis}},\
  }\bibfield  {title} {\enquote {\bibinfo {title} {{Detection of NMR signals
  with a radio-frequency atomic magnetometer}},}\ }\href {\doibase
  http://dx.doi.org/10.1016/j.jmr.2006.12.012} {\bibfield  {journal} {\bibinfo
  {journal} {Journal of Magnetic Resonance}\ }\textbf {\bibinfo {volume}
  {185}},\ \bibinfo {pages} {214 -- 220} (\bibinfo {year} {2007})}\BibitemShut
  {NoStop}%
\end{thebibliography}
\end{document}